\documentclass[preprintnumbers,amsmath,amssymbm,prd]{revtex4}
\usepackage{epsfig}
\usepackage{graphicx}
\usepackage{amssymb}

\begin{document}
\title{How short can stationary charged scalar hair be?}
\author{Shahar Hod}
\affiliation{The Ruppin Academic Center, Emeq Hefer 40250, Israel}
\affiliation{ } \affiliation{The Hadassah Institute, Jerusalem 91010, Israel}
\date{\today}

\begin{abstract}
It is by now well established that charged rotating Kerr-Newman black
holes can support bound-state charged matter configurations which are made of minimally coupled 
massive scalar fields. We here prove that the externally supported stationary charged
scalar configurations {\it cannot} be arbitrarily compact. In particular, for
linearized charged massive scalar fields supported by charged rotating near-extremal
Kerr-Newman black holes, we derive the remarkably compact lower bound
$(r_{\text{field}}-r_+)/(r_+-r_-)>1/s^2$ on the effective lengths of
the external charged scalar `clouds' [here $r_{\text{field}}$
is the radial peak location of the stationary scalar configuration, and
$\{s\equiv J/M^2, r_{\pm}\}$ are respectively the dimensionless
angular momentum and the horizon radii of the central supporting Kerr-Newman black hole].
Remarkably, this lower bound is universal in the sense that it is
independent of the physical parameters (proper mass, electric charge, and angular momentum) of
the supported charged scalar fields.
\end{abstract}
\bigskip
\maketitle

\section {Introduction}

It has recently been proved \cite{Hodrc,HerR} that the composed Einstein-Maxwell-scalar field
theory is characterized by the presence of stationary hairy black-hole solutions with spatially regular horizons. 
The physical and mathematical properties of these hairy Kerr-Newman-massive-charged-scalar-field
configurations have been studied analytically in the linearized regime of the externally supported fields \cite{Hodrc} 
and numerically in the regime of non-linearly coupled massive scalar fields \cite{HerR}.

One remarkable feature of these composed black-hole-charged-massive-scalar-field 
configurations is the fact that they can violate the `no
short hair' lower bound that has been proved in \cite{Hod11} for spherically symmetric hairy 
black-hole spacetimes. In particular, it has
been explicitly shown \cite{Hodnwex} that extremal Kerr-Newman black holes can
support exterior stationary charged scalar fields whose effective
lengths $r_{\text{field}}$ are shorter than the corresponding radius
$r_{\text{null}}$ of the black-hole null circular geodesic:
\begin{equation}\label{Eq1}
r_{\text{field}}<r_{\text{null}}\  .
\end{equation}

The characteristic inequality (\ref{Eq1}) found in \cite{Hodnwex} implies that
the {\it non}-spherically symmetric {\it non}-static composed
Kerr-Newman-charged-massive-scalar-field configurations do not
conform to the no-short-hair lower bound
$r_{\text{field}}>r_{\text{null}}$ proved in \cite{Hod11}, which
states that the externally supported hair of a static spherically-symmetric non-vacuum black hole 
must extend beyond the null circular
geodesic that characterizes the hairy black-hole spacetime.

Motivated by the theorems presented in \cite{Hod11,Hodnwex}, in the present paper 
we raise the following physically intriguing question: How short can 
stationary charged scalar hair be? In particular, one would like to know 
whether the supported charged scalar hair can be made
arbitrarily compact? 

In the present paper we shall provide an explicit answer to this important question. In
particular, below we shall use analytical techniques in order to
derive a remarkably compact lower bound on the effective lengths of the stationary
bound-state linearized charged massive scalar field configurations that are supported in the spacetimes of
near-extremal charged rotating Kerr-Newman black holes.

Interestingly, we shall explicitly show below that the analytically derived 
lower bound [see Eq. (\ref{Eq46}) below] on the effective lengths of the 
charged massive scalar field configurations 
is universal in the sense that it is independent of the physical parameters of
the externally supported scalar fields.

\section{Composed spinning-black-hole-scalar-field configurations}

Early mathematical studies of the composed Einstein-scalar field equations have 
ruled out the existence of spacetime solutions that describe static black-hole-scalar-field hairy
configurations with spatially regular horizons \cite{Chas}. Intriguingly, however, later 
analytical \cite{Hodrc} and numerical \cite{HerR}
studies of the Einstein-matter field equations
have revealed that non-spherically symmetric {\it rotating} black
holes can support stationary matter configurations which are made of spatially regular (neutral or charged) 
massive scalar fields. 

The composed stationary Kerr-Newman-charged-massive-scalar-field configurations
\cite{Hodrc,HerR} that we shall analyze in this paper are intimately related to
the well known physically intriguing phenomenon of superradiant scattering
\cite{Zel,PressTeu1,Bekad} of integer-spin (bosonic) fields in
charged rotating black-hole spacetimes. In particular, the spatially regular 
stationary charged massive scalar field configurations that can be supported 
in the exterior regions of the spinning and charged Kerr-Newman
black-hole spacetime are characterized by orbital frequencies which are in resonance with the threshold
(critical) frequency $\omega_{\text{c}}$ for the superradiant scattering phenomenon 
in the black-hole spacetime \cite{Hodrc,Noteunits}:
\begin{equation}\label{Eq2}
\omega_{\text{field}}=\omega_{\text{c}}\equiv
m\Omega_{\text{H}}+q\Phi_{\text{H}}\ .
\end{equation}
Here the physical parameters $m$ and $q$ \cite{Notedim} are respectively the azimuthal
harmonic index and the charge coupling constant of the supported scalar
field, and \cite{Chan,Kerr,Newman,Notepar,Notesmp}
\begin{equation}\label{Eq3}
\Omega_{\text{H}}={{a}\over{r^2_++a^2}}\ \ \ {\text{and}} \ \ \
\Phi_{\text{H}}={{Qr_+}\over{r^2_++a^2}}
\end{equation}
are the characteristic angular velocity and the electric potential of the central supporting 
Kerr-Newman black hole. The characteristic resonance condition
(\ref{Eq2}) guarantees that the externally supported charged scalar field 
configurations do not radiate their energy and angular momentum into the central supporting charged and
rotating black hole \cite{Hodrc,HerR}.

In addition, the stationary bound-state scalar configurations are
characterized by the inequality \cite{Hodrc,HerR}
\begin{equation}\label{Eq4}
\omega^2_{\text{field}}<\mu^2\  ,
\end{equation}
where $\mu$ \cite{Notedim} is the mass of the supported scalar field. The
characteristic inequality (\ref{Eq4}) guarantees that the stationary massive
scalar field configurations are spatially regular (asymptotically bounded) and that they do 
not radiate their energy to infinity \cite{Hodrc,HerR} [see Eq. (\ref{Eq14}) below].

As emphasized above, former studies \cite{Hodaa,Hod11} of the
Einstein-Maxwell-scalar equations have reveled the physically intriguing fact that charged rotating
Kerr-Newman black holes can support extremely short
bound-state charged scalar field configurations in their exterior regions.
In particular, these non-static non-spherically symmetric
stationary bound-state field configurations were shown
\cite{Hodnwex} to violate the no-short-hair lower bound
$r_{\text{field}}<r_{\text{null}}$ \cite{Hod11} which characterizes
static spherically-symmetric hairy black-hole spacetimes. The main
goal of the present paper is to derive an alternative (and more
robust) lower bound on the effective lengths of these stationary bound-state charged massive linearized 
scalar field configurations. As we shall explicitly show below, for supporting near-extremal
charged rotating Kerr-Newman black holes, the 
no-short-hair lower bound can be derived {\it analytically} [see Eq.
(\ref{Eq46}) below].

\section{Description of the system}

We consider a scalar field $\Psi$ of mass $\mu$ and charge coupling
constant $q$ \cite{Notedim} which is coupled to a
near-extremal Kerr-Newman black hole of mass $M$, angular-momentum
per unit mass $a\equiv J/M$, and electric charge $Q$. 
The stationary scalar fields will be treated at the linear level, 
and we shall therefore use the term `charged scalar clouds' to describe
the externally supported charged massive linearized scalar field configurations
\cite{Noteuh}. As we shall explicitly show below, the main
advantage of the present approach lies in the fact that the linearized
Klein-Gordon wave equation, which determines the spatio-temporal functional behaviors of the composed
Kerr-Newman-black-hole-linearized-charged-massive-scalar-field configurations [see Eq.
(\ref{Eq7}) below], is amenable to an {\it analytical} treatment \cite{Notefn}.

We shall use the Boyer-Lindquist coordinate system, in which case the charged
rotating black-hole spacetime is described by the curved line element
\cite{Chan,Kerr,Newman}
\begin{eqnarray}\label{Eq5}
ds^2=-{{\Delta}\over{\rho^2}}(dt-a\sin^2\theta
d\phi)^2+{{\rho^2}\over{\Delta}}dr^2+\rho^2
d\theta^2+{{\sin^2\theta}\over{\rho^2}}\big[a
dt-(r^2+a^2)d\phi\big]^2\  ,
\end{eqnarray}
where the metric functions are given by the compact expressions $\Delta\equiv r^2-2Mr+a^2+Q^2$ and $\rho^2\equiv
r^2+a^2\cos^2\theta$. The radii of the Kerr-Newman black-hole horizons are given
by the characteristic zeros of of the metric function $\Delta$:
\begin{equation}\label{Eq6}
r_{\pm}=M\pm\sqrt{M^2-a^2+Q^2}\  .
\end{equation}

The temporal and spatial properties of the charged massive linearized scalar field $\Psi$ in the curved black-hole
spacetime are determined by the Klein-Gordon wave equation
\begin{equation}\label{Eq7}
[(\nabla^\nu-iqA^\nu)(\nabla_{\nu}-iqA_{\nu}) -\mu^2]\Psi=0\  ,
\end{equation}
where $A_{\nu}$ is the electromagnetic potential of the charged
black-hole spacetime. It is convenient to use the simple mathematical decomposition
\cite{Noteanz}
\begin{equation}\label{Eq8}
\Psi(t,r,\theta,\phi)=\int\sum_{l,m}e^{im\phi}{S_{lm}}(\theta;m,a\sqrt{\mu^2-\omega^2})
{R_{lm}}(r;M,Q,a,\mu,q,\omega)e^{-i\omega t}d\omega\
\end{equation}
for the scalar eigenfunction $\Psi$, in which case the Klein-Gordon
wave equation (\ref{Eq7}) of the charged massive scalar field in the stationary Kerr-Newman black-hole spacetime (\ref{Eq5}) 
yields two coupled ordinary differential
equations: one equation determines the angular component $S_{lm}$ of the scalar field while
the other equation determines the radial component $R_{lm}$ of the corresponding wave function [see Eqs. (\ref{Eq11}) and (\ref{Eq12}) below].

Using the dimensionless physical parameters \cite{Notedma}
\begin{equation}\label{Eq9}
s\equiv {{a}\over{r_+}}
\end{equation}
and
\begin{equation}\label{Eq10}
\epsilon\equiv \sqrt{\mu^2-\omega^2_{\text{c}}}r_+\  ,
\end{equation}
the angular equation for the scalar eigenfunctions
${S_{lm}}(\theta;m,s\epsilon)$ \cite{Notesa} can be expressed in the
form \cite{Stro,Heun,Fiz1,Teuk,Abram,Hodasy}
\begin{eqnarray}\label{Eq11}
{1\over {\sin\theta}}{{d}\over{\theta}}\Big(\sin\theta {{d
S_{lm}}\over{d\theta}}\Big)
+\Big[K_{lm}+({s}\epsilon)^2\sin^2\theta-{{m^2}\over{\sin^2\theta}}\Big]S_{lm}=0\
.
\end{eqnarray}
The physically motivated requirement of regularity of the angular scalar eigenfunctions
${S_{lm}}(\theta;m,s\epsilon)$ at the two poles, $\theta=0$ and
$\theta=\pi$, singles out a discrete set $\{K_{lm}(s\epsilon)\}$ of
angular eigenvalues (see \cite{Barma,Hodpp} and references
therein).

The radial components ${R_{lm}}$ of the scalar eigenfunctions are
determined by the ordinary differential equation (also known in the physics literature as the radial
Teukolsky equation) \cite{Teuk,Stro}
\begin{equation}\label{Eq12}
\Delta{{d} \over{dr}}\Big(\Delta{{dR_{lm}}\over{dr}}\Big)+\Big\{H^2
+\Delta[2ma\omega-\mu^2(r^2+a^2)-K_{lm}]\Big\}R_{lm}=0\ ,
\end{equation}
where
\begin{equation}\label{Eq13}
H\equiv \omega(r^2+a^2)-ma-qQr\  .
\end{equation}
Note that the eigenvalues $\{{K_{lm}}(s\epsilon)\}$ \cite{Notebr}, which are determined by the angular differential 
equation (\ref{Eq11}), 
also appear in the radial differential equation (\ref{Eq12}) of the charged massive scalar field. 
Thus, these two ordinary differential equations are coupled to each other.

The stationary bound-state configurations of the supported charged massive scalar fields are characterized
by exponentially decaying (bounded) radial eigenfunctions at spatial
infinity \cite{Hodrc,HerR,Ins2}:
\begin{equation}\label{Eq14}
R(r\to\infty)\sim e^{-\epsilon r/r_{\text{H}}}\
\end{equation}
with $\epsilon^2>0$ \cite{Noteepp}. In addition, physically acceptable supported 
field configurations are characterized by spatially regular eigenfunctions. 
In particular, the radial scalar eigenfunctions are finite at the horizon of the 
central supporting black hole \cite{Notepp}:
\begin{equation}\label{Eq15}
0\leq R(r=r_+)<\infty\  .
\end{equation}

\section{The effective radial potential of the composed Kerr-Newman-black-hole-charged-massive-scalar-field configurations}

It proves useful to define the new radial function
\begin{equation}\label{Eq16}
\psi=rR\  ,
\end{equation}
in terms of which the radial equation (\ref{Eq12}) of the charged massive scalar field takes
the form of a Schr\"odinger-like ordinary differential equation
\begin{equation}\label{Eq17}
{{d^2\psi}\over{dy^2}}-V(y)\psi=0\  ,
\end{equation}
where the new radial coordinate $y$ is defined by the relation
\cite{Notemap}
\begin{equation}\label{Eq18}
dy={{r^2}\over{\Delta}}dr\  .
\end{equation}
The effective potential in the differential equation (\ref{Eq17}) is given by the
(rather cumbersome) radial functional expression
\begin{equation}\label{Eq19}
V=V(r;M,Q,a,\mu,q,l,m)={{2\Delta}\over{r^6}}[Mr-(Q^2+a^2)]+{{\Delta}\over{r^4}}
[K_{lm}-2ma\omega_{\text{c}}+\mu^2(r^2+a^2)]-{{1}\over{r^4}}[\omega_{\text{c}}(r^2+a^2)-ma-qQr]^2\
.
\end{equation}

In the next section we shall analyze the near-horizon spatial behavior of the effective
radial potential (\ref{Eq19}) that characterizes the composed
Kerr-Newman-black-hole-charged-massive-scalar-field system. In particular, we
shall prove that $V$ has the form of a potential barrier (namely,
$V\geq0$) in the near-horizon region.

\section{The near-horizon behavior of the radial eigenfunctions}

Defining the dimensionless physical quantities
\begin{equation}\label{Eq20}
x\equiv {{r-r_+}\over{r_+}}\ \ \ \ ; \ \ \ \
\tau\equiv{{r_+-r_-}\over{r_+}}\  ,
\end{equation}
and using the relation (\ref{Eq2}) for the resonant frequency of the supported stationary field, 
one finds the near-horizon behavior
\begin{equation}\label{Eq21}
r^2_+V(x\to0)=F\cdot x(x+\tau)-G\cdot x^2+O(x^3)\
\end{equation}
of the composed Kerr-Newman-charged-massive-scalar-field radial potential (\ref{Eq19}), where the constant ($x$-independent) expansion coefficients are given by
\begin{equation}\label{Eq22}
F\equiv K_{lm}-{{2ma(ma+qQr_+)}\over{r^2_++a^2}}+\mu^2(r^2_++a^2)\ \
\ \ \text{and}\ \ \ \
G\equiv\Big[{{2mar_++qQ(r^2_+-a^2)}\over{r^2_++a^2}}\Big]^2\  .
\end{equation}
Using the lower bound \cite{Barma,Notesi}
\begin{equation}\label{Eq23}
K_{lm}\geq m^2-a^2(\mu^2-\omega^2_c)\
\end{equation}
on the eigenvalues of the angular equation (\ref{Eq11}), and taking cognizance of
the inequality (\ref{Eq4}), which characterizes the bound-state resonances of the composed black-hole-massive-scalar-field system, 
one finds the characteristic inequality
\begin{equation}\label{Eq24}
F>[m^2+(qQ)^2]{{r^2_+}\over{r^2_++a^2}}>0\  .
\end{equation}

In the near-horizon region,
\begin{equation}\label{Eq25}
x\ll\tau\  ,
\end{equation}
one finds the relation [see Eqs. (\ref{Eq18}) and (\ref{Eq20})]
\begin{equation}\label{Eq26}
y={{r_+}\over{\tau}}\ln(x)+O(x)\  ,
\end{equation}
which implies \cite{Noteyas}
\begin{equation}\label{Eq27}
x=e^{\tau y/r_+}[1+O(e^{\tau y/r_+})].
\end{equation}
Taking cognizance of Eqs. (\ref{Eq17}), (\ref{Eq21}), and
(\ref{Eq27}), one obtains the Schr\"odinger-like wave equation
\begin{equation}\label{Eq28}
{{d^2\psi}\over{d\tilde y^2}}-{{4F}\over{\tau}}e^{2\tilde y}\psi=0\
\end{equation}
in the near-horizon region (\ref{Eq25}), where
\begin{equation}\label{Eq29}
\tilde y\equiv {{\tau}\over{2r_+}}y\  .
\end{equation}
The mathematical solution of the near-horizon radial differential equation (\ref{Eq28}) that
respects the physically motivated boundary condition (\ref{Eq15}) can be expressed in a compact form in terms of the 
modified Bessel function of the first kind \cite{Noteab1,Notesk}:
\begin{equation}\label{Eq30}
\psi(y)=I_0\Big(2\sqrt{{{F}\over{\tau}}}e^{\tau y/2r_+}\Big)\  .
\end{equation}
Using the well-known properties of the modified Bessel function
$I_0$ \cite{Abram}, one deduces from (\ref{Eq30}) that $\psi(y)$ is
a positive, increasing, and convex function in the near-horizon
$x\ll\tau$ region. That is,
\begin{equation}\label{Eq31}
\{\psi>0\ \ \ ;\ \ \ {{d\psi}\over{dy}}>0\ \ \ ;\ \ \
{{d^2\psi}\over{dy^2}}>0\}\ \ \ \ \text{for}\ \ \ \ 0<x\ll\tau\ .
\end{equation}

We shall now prove that the radial function $\psi[y(x)]$, which characterizes the spatial behavior 
of the charged massive scalar fields in the charged and spinning black-hole spacetime, is a positive, increasing, and
convex function in the finite radial interval $[0,x_o]$, where the location
of the outer boundary is given by
\begin{equation}\label{Eq32}
{{x_o}\over{\tau}}={{F}\over{G-F}}\  .
\end{equation}
To this end, we first note that from Eqs. (\ref{Eq22}) and
(\ref{Eq24}) one learns that the effective near-horizon radial
potential (\ref{Eq21}) has the form of a potential barrier. In
particular, one finds
\begin{equation}\label{Eq33}
V\geq0\ \ \ \text{for}\ \ \ x\in [0,x_o]\  ,
\end{equation}
where the outer turning point $x=x_o$ of the near-horizon potential
barrier (\ref{Eq21}) is given by the relation (\ref{Eq32})
\cite{Noteoi,Noteinp}. From the radial differential equation (\ref{Eq17}) and the inequality (\ref{Eq33}) one
deduces that $\psi[y(x)]$ is a convex function in the entire
interval $(0,x_o)$. That is,
\begin{equation}\label{Eq34}
{{d^2\psi}\over{dy^2}}>0\ \ \ \ \text{for}\ \ \ \ x\in (0,x_o)\ .
\end{equation}

The relation (\ref{Eq34}) implies that $d\psi/dy$ is an increasing
function in the entire interval $(0,x_o)$. Remembering that
$d\psi/dy>0$ in the near-horizon $x\ll\tau\ (y\to\-\infty)$ region
[see Eq. (\ref{Eq31})], one deduces that $\psi[y(x)]$ is a monotonically 
increasing function in the entire radial interval $(0,x_o)$. That is,
\begin{equation}\label{Eq35}
{{d\psi}\over{dy}}>0\ \ \ \ \text{for}\ \ \ \ x\in (0,x_o)\ .
\end{equation}
Furthermore, remembering that $\psi>0$ in the near-horizon
$x\ll\tau\ (y\to\-\infty)$ region [see Eq. (\ref{Eq31})], one
deduces from (\ref{Eq35}) that the radial scalar eigenfunction $\psi[y(x)]$ is a positive definite
function in the entire interval $(0,x_o)$. That is,
\begin{equation}\label{Eq36}
\psi>0\ \ \ \ \text{for}\ \ \ \ x\in (0,x_o)\ .
\end{equation}

Taking cognizance of Eqs. (\ref{Eq34}), (\ref{Eq35}), and
(\ref{Eq36}), one finally finds that $\psi[y(x)]$ is a positive,
increasing, and convex function in the interval $(0,x_o)$. That is,
\begin{equation}\label{Eq37}
\{\psi>0\ \ \ ;\ \ \ {{d\psi}\over{dy}}>0\ \ \ ;\ \ \
{{d^2\psi}\over{dy^2}}>0\}\ \ \ \ \text{for}\ \ \ \ x\in (0,x_o)\ ,
\end{equation}
where the value of the outer radial point $x_o$ is given by the compact expression (\ref{Eq32}). The relations
(\ref{Eq37}) imply, in particular, that the radial function
$\psi(x)$, which characterizes the spatial behavior of the supported charged massive scalar fields in the charged and spinning 
black-hole spacetime, has no local maximum points within the radial 
interval $(0,x_o)$.

Moreover, the fact that $\psi(x)$ is a positive increasing function
in the interval $x\in (0,x_o)$ [see Eq. (\ref{Eq37})], together with
the asymptotic boundary condition (\ref{Eq14}), which characterizes the bound-state resonances of 
the composed black-hole-massive-scalar-field system, imply that $\psi(x)$
has (at least) one maximum point, $x=x_{\text{peak}}$, which is
located {\it outside} this interval. That is,
\begin{equation}\label{Eq38}
x_{\text{peak}}\geq x_o\  .
\end{equation}
In the next section we shall obtain a generic lower bound on the radial 
peak location $x_{\text{peak}}$ of the supported scalar eigenfunction
$\psi(x)$.

\section{A lower bound on the effective lengths of the supported charged-massive-scalar-field configurations}

In the present section we shall derive a remarkably compact lower bound on the value of
the outer turning point $x_o$ [see Eq. (\ref{Eq32})] of the
effective radial potential (\ref{Eq21}). This, in turn, would yield a lower bound on the peak location, $x_{\text{max}}$, of the radial
eigenfunctions $\psi(x)$ that characterize the composed
Kerr-Newman-black-hole-charged-massive-scalar-field configurations.

Substituting (\ref{Eq22}) and (\ref{Eq24}) into (\ref{Eq32}), we
find the parameter-dependent lower bound \cite{NoteFG}
\begin{equation}\label{Eq39}
{{x_o}\over{\tau}}>{\cal R}(s,\gamma)
\end{equation}
on the location of the outer radial turning point, where
\begin{equation}\label{Eq40}
{\cal R}(s,\gamma)\equiv
{{s^2+\gamma^2}\over{3s^2-1+4\gamma(1-s^2)-\gamma^2(3-s^2)}}\cdot{{1+s^2}\over{s^2}}
\end{equation}
and
\begin{equation}\label{Eq41}
\gamma\equiv {{qQs}\over{m}}\  .
\end{equation}

As explained above, our main goal in this paper is to derive, using analytical techniques, 
a {\it generic} lower bound \cite{Notegen} on the effective lengths (radii)
of the externally supported stationary charged massive scalar field configurations. Hence, we
shall now determine the particular value of the charge coupling parameter 
$q=q_{\text{min}}$ which, for given physical parameters $\{M,Q,a\}$ of the
central supporting Kerr-Newman black hole, {\it minimizes} the dimensionless
function ${\cal R}(s,\gamma)$ in the lower bound (\ref{Eq39}).

Differentiating ${\cal R}(s,\gamma)$ with respect to $\gamma$, one
finds that ${\cal R}(s,\gamma)$ is minimized for
\begin{equation}\label{Eq42}
\gamma^*(s)={{1-s^2}\over{2}}\  .
\end{equation}
Substituting (\ref{Eq42}) back into the expression (\ref{Eq40}), one obtains
\begin{equation}\label{Eq43}
{\cal R}^{*}(s)\equiv\text{min}_{\gamma}\{{\cal
R}(\gamma;s)\}={{1}\over{s^2}}\ .
\end{equation}
Taking cognizance of Eqs. (\ref{Eq39}) and (\ref{Eq43}), one finds
the lower bound
\begin{equation}\label{Eq44}
{{x_o}\over{\tau}}>{{1}\over{s^2}}\geq1\
\end{equation}
on the location of the outer radial turning point. 

Before proceeding, we would like to stress the fact that we have kept terms of order
$O(x^2,x\tau)$ in the near-horizon expansion (\ref{Eq21}) of the
effective radial potential but neglected terms of order $O(x^3)$.
The requirement $x^3_o\ll x_o\tau$ together with the inequality
$x_o>\tau/s^2$ [see Eq. (\ref{Eq44})] imply that our analysis is
self-consistent in the near-extremal regime \cite{Notelta}
\begin{equation}\label{Eq45}
\tau\ll s^4\  .
\end{equation}

Finally, taking cognizance of the inequality (\ref{Eq38}), we obtain
the {\it generic} (that is, {\it independent} of the field
parameters) lower bound
\begin{equation}\label{Eq46}
{{x_{\text{peak}}}\over{\tau}}>{{1}\over{s^2}}\geq1\
\end{equation}
on the peak location of the radial eigenfunctions $\psi(x)$ that
characterize the composed Kerr-Newman-black-hole-charged-massive-scalar-field
configurations.

\section{Summary}

The `no short hair' theorem presented in \cite{Hod11} has revealed the physically intriguing fact that
spherically-symmetric static hairy black-hole configurations cannot
be arbitrarily compact. In particular, the theorem proved in
\cite{Hod11} asserts that the external fields of a
spherically-symmetric static hairy black hole must extend beyond the
null circular geodesic of the black-hole spacetime.

Interestingly, it has been demonstrated \cite{Hodnwex} that non-static 
non-spherically symmetric hairy black-hole configurations may 
violate this lower bound. In particular, it has been explicitly shown in
\cite{Hodnwex} that extremal charged rotating Kerr-Newman black
holes can support linearized charged scalar field configurations (stationary charged scalar `clouds') 
whose effective lengths are characterized by the inequality $r_{\text{field}}<r_{\text{null}}$.

Motivated by the intriguing finding presented in \cite{Hodnwex}, that non-static non-spherically symmetric 
composed black-hole-charged-scalar-field configurations can violate the 
lower bound $r_{\text{field}}>r_{\text{null}}$ \cite{Hod11} which characterizes 
static spherically-symmetric hairy black-hole spacetimes \cite{Notepn}, 
in the present paper we have raised 
the following physically interesting question: How short can stationary charged scalar hair be?

In order to address this physically important question, we have studied, using analytical techniques, the
Klein-Gordon-Kerr-Newman wave equation for a stationary charged massive scalar
field which is linearly coupled to a near-extremal charged rotating
Kerr-Newman black hole. Interestingly, it has been explicitly shown that the externally supported stationary charged scalar
configurations cannot be made arbitrarily compact. In particular, we have proved analytically that the radial peak location
$r_{\text{peak}}$ of the scalar eigenfunction $\psi(r)$, which
characterizes the composed Kerr-Newman-black-hole-charged-massive-scalar-field
configurations, is bounded from below by the remarkably compact functional relation [see Eqs.
(\ref{Eq20}) and (\ref{Eq46})]
\begin{equation}\label{Eq47}
{{r_{\text{peak}}-r_+}\over{r_+-r_-}}>{{1}\over{s^2}}
\  .
\end{equation}
It is physically interesting to emphasize the fact that the analytically derived lower bound (\ref{Eq47}) 
is universal in the sense that it is {\it independent} of the physical parameters 
(proper mass, electric charge, and angular harmonic indexes) of the supported scalar fields.

\bigskip
\noindent
{\bf ACKNOWLEDGMENTS}
\bigskip

This research is supported by the Carmel Science Foundation. I thank
Yael Oren, Arbel M. Ongo, Ayelet B. Lata, and Alona B. Tea for
stimulating discussions.



\begin{thebibliography}{99}
\bibitem{Noteunits} We use natural units in which $G=c=\hbar=1$.
\bibitem{Notedim} Note that $\mu$ and $q$, the mass and charge parameters of the externally supported scalar fields, 
stand respectively for $\mu/\hbar$ and $q/\hbar$. Hence, these characteristic 
physical parameters of the spatially regular charged massive scalar fields have the dimensions of $($length$)^{-1}$.
\bibitem{Notepar} Here $a, Q$, and $r_+$ are respectively the angular momentum per unit mass,
electric charge, and horizon-radius of the charged rotating
Kerr-Newman black-hole spacetime.
\bibitem{Notesmp} We shall henceforth assume $a>0, Q>0$, and $m>0$ without loss of
generality.
\bibitem{Noteuh} Note that the term `scalar hair' is usually used in the physics literature to
describe externally supported spatially regular scalar fields which are 
non-linearly coupled to the central black hole.
\bibitem{Notefn} 
We believe that it would be highly interesting to use more sophisticated numerical
techniques \cite{HerR} in order to generalize our analytical results to the
regime of stationary bound-state charged massive scalar field configurations that are 
non-linearly coupled to the central supporting charged and spinning black holes.
\bibitem{Noteanz} The physical parameters $\omega,\ m$, and $l\geq|m|$ are respectively the conserved
frequency and the harmonic indexes (azimuthal and spheroidal) of the linearized scalar
field mode [see Eq. (\ref{Eq11}) below].
\bibitem{Notedma} The physical parameter $s\equiv J/Mr_+$ is the dimensionless angular-momentum of the 
charged and spinning central supporting Kerr-Newman black hole.
\bibitem{Notesa} The functions ${S_{lm}}(\theta;m,s\epsilon)$, which characterize the angular behavior of the 
scalar fields in the spinning black-hole spacetime, are known in the physics literature as the spheroidal harmonic eigenfunctions.
\bibitem{Notebr} For brevity, we shall henceforth omit the dimensionless angular harmonic indexes $\{l,m\}$ that characterize 
the eigenfunctions of the supported charged massive scalar fields.
\bibitem{Noteepp} One may choose $\epsilon>0$ without loss of generality.
\bibitem{Notepp} One can assume $R(r=r_+)\geq0$ without loss of generality.
\bibitem{Notemap} Note that the radial transformation (\ref{Eq18}) maps the semi-infinite Boyer-Lindquist 
radial coordinate $r\in [0,\infty]$ into the new (infinite) radial coordinate $y\in [-\infty,+\infty]$.
\bibitem{Notesi} Note that the lower bound (\ref{Eq23}) on the eigenvalues of the angular scalar equation (\ref{Eq11}) 
can be approached in
the eikonal $l=m\gg1$ regime, in which case one finds the simple relation
$K_{mm}(s\epsilon)=m^2[1+O(m^{-1})]-({s}\epsilon)^2$ \cite{Hodpp}.
\bibitem{Noteyas} Note that the near-horizon region (\ref{Eq25}) corresponds to the relation $y\to
-\infty$ [see Eq. (\ref{Eq26})], which yields the simple limit $e^{\tau y/r_+}\to
0$.
\bibitem{Noteab1} Here we have used Eqs. (9.1.54) and (9.6.3) of \cite{Abram}. Note that the normalization 
constant of the linearized solution (\ref{Eq30}) can be chosen, without loss of generality, to be $1$.
\bibitem{Notesk} The second mathematical solution of the near-horizon radial 
differential equation (\ref{Eq28}) is given by the modified
Bessel function of the second kind:
$\psi_2(y)=K_0(2\sqrt{{{F}/{\tau}}}e^{\tau y/2r_+})$. However, using
Eq. (9.6.13) of \cite{Abram}, one finds that this solution does not
respect the physically motivated boundary condition (\ref{Eq15}) at the black-hole
horizon (In particular, one finds the divergent behavior $\psi_2(y\to -\infty)\to -\tau
y/2r_+\to\infty$ at the black-hole horizon).
\bibitem{Noteoi} Note that if $G-F\leq0$ then the
outer turning point of the effective radial potential is characterized by the relation
$x_o/\tau\gg1$. This implies that the function $\psi(x)$ has a maximum at some
$x=x_{\text{peak}}\geq x_o\gg\tau$ [see Eq. (\ref{Eq38}) below]. 
Thus, in this case we recover our final conclusion [see Eq.
(\ref{Eq46}) below] that the non-trivial behavior of the externally supported 
scalar fields must extend beyond $x=\tau$.
\bibitem{Noteinp} Note that the inner turning point of the effective potential barrier is trivially given by
$r=r_+$ (that is, $x=0$) [see Eq. (\ref{Eq21})].
\bibitem{NoteFG} Here we have used the relations $x_{\text{o}}/\tau=F/(G-F)>F/G$ [see Eqs. (\ref{Eq24}) and (\ref{Eq32})].
\bibitem{Notelta} Note that the 
requirement $x^3_o\ll x^2_o$ yields the weaker constraint $\tau\ll s^2$.
\bibitem{Notepn} We would like to emphasize again that the no-short hair lower bound
$r_{\text{field}}>r_{\text{null}}$ was rigorously proved in
\cite{Hod11}, using analytical techniques, for the case of static spherically-symmetric hairy black-hole spacetimes.
\end{thebibliography}
\end{document}